\documentstyle[sprocl,psfig]{article}

\begin{document}

\title{GENERAL INTRODUCTION TO COMPTON SCATTERING}

\author{D. DRECHSEL}

\address{Institut fuer Kernphysik, Johannes Gutenberg-Universitaet Mainz,\\ 
Staudinger Weg 7, D-55099 Mainz}

\maketitle\abstracts{Real and virtual Compton scattering has been and will be an important tool to study the structure of hadronic systems. We summarize the status of real Compton scattering and give an outlook at the new theoretical 
and experimental developments in the field of virtual Compton scattering.}

\section{The Classical Concept of Polarizabilities}

The polarizability of a system defines its response to a quasistatic electromagnetic field. The induced dipole moment of two constituents of opposite charge, $q$ and $-q$ respectively, at a distance $\vec{r}$ is $\vec{d} = q \vec{r} =
\alpha \vec{E}$,
with $\alpha$ the electric polarizability and $\vec{E}$ the electric field. If the force between the constituents is governed by Hooke's law, the equilibrium is obtained for $\alpha = 2q^{2}/C$ where $C$ is the spring constant. 
In the presence of an electromagnetic field, the total energy of the system 
decreases by $\Delta E = - \frac{1}{2} \alpha \vec{E}^{2}$.

The speed of light in the medium, $v$, is lowered by the polarization 
of the medium, and the index of refraction, $n$, changes correspondingly. As 
a consequence light is scattered, the forward scattering amplitude being proportional to the square of the polarizabilities. 

In the case of a metal sphere in a homogeneous electrical field, $\alpha\simeq
R^{3}$, where $R$ is the radius of the sphere. For a dielectric sphere, the 
polarizability is very much reduced, $\alpha/R^3 = (\epsilon -1)/(\epsilon 
+ 2)$, with $\epsilon \approx 1$ for most media. The same is the case for the nucleon, where $\alpha_{N}/R_{N}^{3} \approx 10^{-3}$. In conclusion, the nucleon is a very rigid object, with strong ''springs'' between its constituents.

Similar models apply for the magnetic polarizability (or susceptibility) $\beta$. In that case we expect both the orientation of preformed magnetic moments
along the direction of the exterior magnetic field, leading to an enhancement of the total field $\beta_{para}$ (paramagnetism), and induced electric currents leading to induced magnetization $\beta_{dia}$, weakening the field by Lenz's law (diamagnetism). 

\section{Compton Scattering}

The general kinematics for Compton scattering ($CS$) is shown in Figure 1. 
In the case of real Compton scattering ($RCS$), the squares of the four-
momenta are $
q^2 = q^{'2} = 0$. Electron scattering 
leads to $Q^2 = - q^{2} > 0,$ e.g. virtual photons with space-like 
properties. By observing electron-positron pairs in the final state, we could select reactions with time-like virtual photons, $Q^{'2} = - q^{'2}<0$
\cite{Lvo96}. These reactions with at least one virtual photon are called virtual Compton
scattering ($VCS$). Being of purely electromagnetic origin, they 
probe the charge, current and magnetization of the quarks in the hadronic system. In that sense $RCS$ and $VCS$ are important complements to measurements 
of form factors or deep inelastic scattering (DIS) off the
constituents of hadrons.
\begin{figure}
\centerline{\psfig{figure=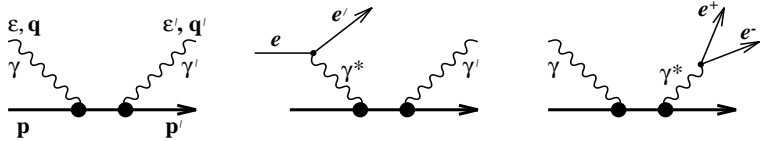,width=10cm}}
\caption{Kinematics for Compton scattering. The incoming (real or virtual) 
photon has polarization $\epsilon$ and momentum $q = (\omega, \vec{q})$, the momentum of the hadron in the initial state is $p$. The corresponding variables in the final state appear with a prime. The Mandelstam variables are 
$s = (p+q)^{2}, t = (q' - q)^{2}$ and $u = (p - q')^{2}$.
\label{fig1}}
\end{figure}
There are three regions of particular physical interest in $CS$:

(I) $\bf{Threshold}$: For $\omega_{cm}\le m_{\pi}$, the scattering amplitudes are real except for small electromagnetic corrections. The model-dependence of the cross-section enters essentially by the polarizabilities, which give global information on the excitation spectrum of the hadron \cite{Lvo93a}.

(II) $\bf{Resonances}$: In the region of the low-lying resonances, 
$\Delta$(1232), $N^{\ast}$ (1440, 1520 and 1535), $CS$ provides independent information on the resonance multipoles and an excellent check on the 
consistency of the analysis. At the higher energies it has been suggested to look for ''missing resonances'' \cite{And93}.

(III) $\bf{High\,energies}$: In the limit of Mandelstam variable 
$s\rightarrow\infty$ perturbative $QCD$ is supposed to work.  It 
predicts a factorization of the transition amplitudes into a distribution
amplitude (''wave function'') and an elementary scattering amplitude 
off current quarks \cite{Far90}. The Fock state of 3 valence quarks should give
the leading contribution to the cross section, which is expected to scale
like $s^{-6}$ at constant $t$. The higher Fock states should be
suppressed by factors of $Q^{-2}$ for each additional $q\bar{q}$
pair. 

The general amplitude for $CS$ must be linear in the polarization vectors of the initial and final photon, $T = \epsilon'^{\mu} \epsilon^{\nu} T_{\mu\nu}$,
with the Compton tensor $T_{\mu\nu}$ to be constructed from the independent momenta and the Dirac matrices of the nucleon. Using Lorentz invariance, gauge
invariance and crossing symmetry ($s\leftrightarrow u$), we find in the case of $RCS$ that there are 6 independent structures $O_{i}$ \cite{Lea62,Rol76}, 
which appear with 6 Lorentz invariant amplitudes  $A_{i}$,\\
\begin{equation}
T = \sum\limits^{6}_{i = 1} A_{i} (s, t) O_{i}\,\,. 
\end{equation}
Below pion production threshold, $s \le (m + m_{\pi})^{2}$, these amplitudes
are essentially real. Two of the amplitudes do not depend on the spin of the
nucleon, they are related to the scalar polarizabilities $\alpha$ and $\beta$ introduced before. The matrix elements of the relativistic invariants $O_i$ between the nuclear spinors may be expressed in the usual form $a 1 + i\vec{b}\cdot
\vec{\sigma}$. As an example the forward scattering amplitude obtains the 
form
\begin{eqnarray}
4 \pi f (\omega, \Theta = 0) & = & \hat{\epsilon}'\cdot \hat{\epsilon}
f_{1} (\omega) + i \omega \vec{\sigma}\cdot (\hat{\epsilon}'\times
\hat{\epsilon}) f_{2} (\omega),\nonumber\\
f_{1} (\omega) & = & - \frac{e_{N}^{2}}{m} + (\alpha + \beta) \omega^{2} 
+ [\omega^{4}]\\
f_{2} (\omega) & = & - \frac{e^{2} \kappa_{N}^{2}}{8\pi m^{2}} + \gamma
\omega^{2} + [\omega^{4}]\,\,. \nonumber
\end{eqnarray}
The leading contribution in the limit $\omega\rightarrow 0 $ is the Thomson
amplitude depending only on global properties of the nucleon, its charge 
$e_{N} (+ e$ for the proton, 0 for the neutron) and its mass $m$. 
We note that the spin-independent forward amplitude $f_{1}$ contains only the 
sum of $\alpha$ and $\beta$. In order to obtain the two quantities 
separately, the full angular dependence has to be studied. By averaging the 
square of the scattering amplitude over the polarization of the
photon, one finds that the cross section at $\Theta = 90^{\circ}$ contains only
$\alpha$, while at $\Theta = 180^{\circ}$ the difference ($\alpha - \beta$) may be determined.
We further note that the leading term of the Born amplitude (point nucleon with anomalous magnetic moment $\kappa_{N}$) yields the Thomson limit, 
$d\sigma/d\Omega = \frac{1}{2} (1+\cos^{2} \Theta) r_{0}^{2}$, with $r_0^{2} = (e^{2}/m)^2 
\approx 24 nb/s r$.

The spin-dependent amplitude $f_2$ contains as leading term the anomalous
magnetic moment $\kappa_{N}$ and as next-to-leading term one of the 
spin-dependent or vector polarizabilities $\gamma$. The measurement of all 4 spin-dependent polarizabilities would require to study the angular distribution for polarized nucleons and photons. 

In the case of $VCS$ the number of independent amplitudes increases as shown
in Table 1 \cite{Ber61,Ber76}.
By comparing $RCS$ and $VCS$, we see that there are 
two transverse spin-independent polarizabilities and an additional longitudinal one appearing in the case of $VCS$. Of the 9 spin-dependent polarizabilities
of $VCS$, the 3 longitudinal ones cannot be observed in the case of $RCS$. Of the remaining 6 transverse ones only 4 remain in the case of $RCS$ because of time reversal. Obviously, $VCS$ is much richer in information than $RCS$. Whether this potential can be fully exploited remains to be seen. 

%\subsection{}\label{}
%\subsection{}
%\subsection{Tables}
\begin{table}[t]
%\begin{flushleft}
\caption{Number of independent amplitudes for $RCS(\gamma\gamma), VCS (\nu^{\ast}\nu)$ and doubly virtual $CS(\nu^{\ast}\nu^{\ast}$. The
second column ($1$) denotes the spin-independent, the third column 
($\vec{\sigma}$) the spin-dependent amplitudes}

%\end{flushleft}
%\vspace{0.4cm}
\begin{center}
\begin{tabular}{|c|c|c|c|c|}
%\hline
\hline
 & $1$& $\vec{\sigma}$ & sum & Ref. \\
\hline
$\gamma\gamma$ & 2 & 4 & 6 & [5,6] \\
\hline
$\gamma^{\ast}\gamma$ & 3 & 9 & 12 & [7,8]\\
\hline
$\gamma^{\ast}\gamma^{\ast}$ & 5 & 13 & 18 & [9] \\
\hline
\end{tabular}
\end{center}
\end{table}
\section{Low Energy Theorem}

Due to crossing symmetry, $f_1$ and $f_2$ of equation (2) are even functions 
of $\omega = \omega_{cm}$. The value of both functions at $\omega = 0$ is determined by global properties of the hadronic system (total charge, mass, magnetic moment), and only the terms O($\omega^{2}$) give new information depending on 
the interior degrees of freedom (polarizabilities). This behaviour has been
established by the low energy theorem ($LET$) due to Gell-Mann, Goldberger 
\cite{Gel54} and Low \cite{Low54}. By inspecting Figure 1 we find that the propagator of the intermediate state has the energy dependence ($s - M^{2})^{-1}$, where $M$ is the mass of the intermediate state. Since 
$s = (p+q)^2 = m^2 + 2p\cdot q + q^{2}$, the mass terms 
cancel for an intermediate nucleon ($ M = m$), leading to a 
propagator with an infrared divergence $O (q^{-1})$. For excited states the infrared limit is $(m^2 - M^2)^{-1}$ , where $M = m_{\ast}$ for nucleon resonances
or $M > m + m_{\pi}$ for pion loops etc. Hence intermediate states different
from the ground state (nucleon) do not give rise to divergencies but are 
$O (1)$. Therefore, the Compton amplitude can be written in the form
$T = T_{Born} + T_{h.o.t}$, 
with $T_{Born} = O (q^{-1})$ including all the tree-level diagrams with nucleon poles in the $s$- or $u$-channel (crossed diagram), pion poles in the $t$-channel and, if necessary, contact terms (see Figure 2). 
All higher order terms (resonances, loops) are contained in the second term which does not lead to infrared 
divergencies in the limit $q \rightarrow 0$. It is the essence of $LET$ that also the terms $O (1)$ can be determined in a model-independent way due to gauge invariance.
\begin{figure}
%\rule{5cm}{0.2mm}\hfill\rule{5cm}{0.2mm}
%\vskip 2.5cm
%\rule{5cm}{0.2mm}\hfill\rule{5cm}{0.2mm}
\centerline{\psfig{figure=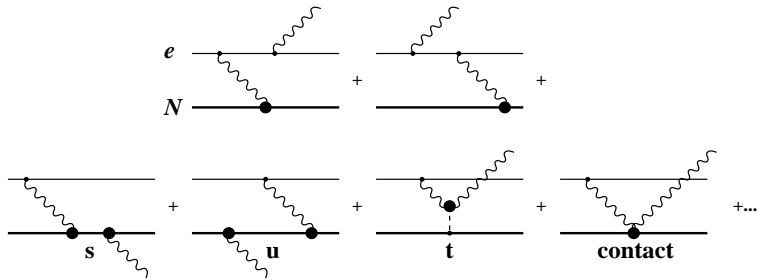,width=10cm}}
\caption{Leading order Feynman diagrams contributing to Bethe-Heitler radiation
(top) and $VCS$ (bottom)
\label{fig2}}
\end{figure}
The exact proof of that statement requires to write down the most general form of the Compton tensor $T_{\mu\nu}$ expressed in terms of the independent combinations of momenta and Dirac matrices, expanded as a power series in $q$ and 
$q'$. A qualitative understanding is easily obtained for a particle without 
spin (e.g. a pion), for which the tensor structure must be built up by bilinear forms of the independent momenta $q, q'$ and $P = \frac{1}{2} (p + p')$. 
In a somewhat symbolic form we have 
\begin{equation}
T^{\mu\nu} = (\frac{a_{s}}{p\cdot q} + \frac{a_{u}}{p'\cdot q} + a_{c} + ...) 
P^{\mu} P^{\nu} + b P^{\mu} P^{\nu} + O (q, q')\,,\nonumber
\end{equation}
where the first terms on the $r.h.s.$ describe the $s$- and $u$-channel
singularities and a contact term at tree-level, and the second term contains
the lowest order contribution, $O (1)$, of all higher order (loop)  diagrams. 
Gauge invariance requires $ q_{\mu} T^{\mu\nu} = T^{\mu \nu} q'_{\nu}$,
and if the tree level (Born) diagrams are gauge-invariant by themselves 
(as they should be for a gauge-invariant Lagrangian !), we find $b = 0$ , 
because  $q\cdot P P^{\nu} \not= 0$ in general. Hence the (unknown)  loop diagrams are contained in the terms $O (q, q')$. For the careful reader we add that the argument does not change if the Born terms contain additional tensor structures of the type $P^{\mu}q^{\nu}, P^{\mu}q'^{\nu}$ or  $q^{\mu}
q'^{\nu}$, because we require that the complete Born amplitude should fulfil gauge invariance. In any case 
the leading term of the non-Born amplitude must be of the structure 
$P^{\mu}P^{\nu}$, because any
additional $q$ or $q'$ would necessarily make it a term of $O (q, q')$. 
We summarize that the leading two powers in $q$ or $q'$ are determined by the Born amplitude, and this is the reason why the model-dependence of the functions $f_1$ and $f_2$ arises only at $O (\omega^{2})$.

\section{Dispersion Relations}

Dispersion relations give a connection between the photon scattering amplitude and the various reaction channels. They are based on analyticity, unitarity and crossing symmetry. Analyticity implies that the real and imaginary parts of the scattering amplitude are Hilbert transforms of each other.
Using crossing symmetry and analyticity, the integral over the negative values of $\omega$ can be expressed by the positive ones. If the resulting
 integral does not yet converge, it will be necessary to subtract the 
function at the origin.

The spin-independent forward scattering amplitude can be easily evaluated, 
because unitarity provides the optical theorem $ Im f (\omega) = 
\omega \sigma (\omega)/4 \pi$,
where $\sigma (\omega)$ is the total absorption cross section. As a consequence
\begin{equation}
Re f_{1}(\omega) = f_{1}(0) + \frac{\omega^{2}}{2\pi^{2}} \int d\omega'
\frac{\sigma (\omega ')}{\omega^{'2} - \omega^{2}}\,.
\end{equation}
By comparing this relation with equation (2) for small $\omega^{2}$ we find 
Baldin's sum rule \cite{Bal60},
\begin{equation}
\alpha + \beta = \frac{1}{2\pi^{2}} \int\limits^{\infty}_{thr} d\omega
\frac{\sigma (\omega)}{\omega^{2}} = (14.2 \pm 0.5) 10^{-4} fm^{3}\,\,. 
\nonumber
\end{equation}
A similar argument can be used for the spin-dependent amplitude. Defining the 
amplitudes $f_{3/2}$ and $f_{1/2}$ for parallel and antiparallel spins of photon and nucleon,we have $4\pi f_{\{1/2, 3/2\}} = f_{1} \pm \omega f_{2}$,
with the plus or minus sign for $f_{1/2}$ and $f_{3/2}$, respectively. The optical theorem relates the imaginary part of the amplitudes to the absorption cross section in the corresponding spin channels, and
similar arguments as above lead to 
\begin{equation}
I (0) = \frac{1}{2\pi^{2}} \int\limits^{\infty}_{thr} d\omega 
\frac{\sigma_{1/2} (\omega, 0) - \sigma_{3/2} (\omega, 0)}{\omega}
= - \frac{e^{2} \kappa^{2}}{4\pi m^{2}}\,\,.\nonumber
\end{equation}
This is the Gerasimov-Drell-Hearn sum rule \cite{Ger66,Dre66} relating the anomalous magnetic moment to the difference of the absorption cross sections in the two spin channels. We note in particular that the integral must be larger for spin $3/2$ than for spin $1/2$ resonances. As always it is assumed that the integral exists. It should also be pointed out that the GDH sum rule is the real photon limit of a more general integral $ I(Q^{2})$, which  has been measured in $DIS$ off the nucleon. In the region $Q^{2}\ge 1 GeV^{2}$ the integral has a (small) positive value, while it is obviously negative at the photon point. The cross-over from negative to positive values is expected at $Q^{2} \approx$
$m_{V}^{2}$, in the region of the vector meson masses. Since $DIS$ has led to 
interesting results about the spin structure of the nucleon (''spin crisis''), the expected rapid increase of the integral for $Q^{2}< 1 GeV^{2}$ is of considerable theoretical interest and will be investigated in future experiments at 
CEBAF \cite{Bur91,Kuh93}. 

Dispersion relations at finite scattering angle are more delicate. They have to be set up for the invariant amplitudes $A_i$.
Unfortunately, the dispersion integrals for the amplitudes $A_1$ and $A_2$ are not expected to converge because of a Regge type behaviour at high energies,
$A_{1,2} \sim \omega^{\alpha_{1,2}(t)}$. Clearly further subtraction at small $\omega$ would improve the situation, but the new subtraction functions would include the polarizabilities. In order to keep a predictive power for the low energy amplitude, L'vov \cite{Lvo93b} has proposed to perform the integral on the real axis only up to $\omega_{max} = 1.5 GeV$ and then to close the contour by a large half-circle around the origin. The main contribution on the circle is  assumed to come from $t$- channel exchange of the $\pi^{0}$(triangle anomaly) in the case of $A_2$ and of $\pi\pi$ states with the  quantum number of the fictitious $\sigma$-meson for $A_1$. This introduces, of course, some model-dependence in $A_1$ and $A_2$, while the remaining amplitudes $A_3 - A_6$ converge reasonably well. In particular we find $2\pi (\alpha - \beta) = - A_{1}$ and $2\pi (\alpha + \beta) = 
- A_{3} - A_{6}$,
i.e. the sum of the polarizabilities is well under control, while the difference is largely influenced by high energy contributions. Table 2 shows the 
decomposition of $\alpha$ and $\beta$ into the contributions of the integrals
along the real axis and the half-circle. It is obvious that at least half of the electric polarizability comes from the high energy region. The table also explains why $\beta$  is so small. The large paramagnetic contribution, mostly from the $\Delta (1232)$, is cancelled by diamagnetic contributions essentially due to $t$-channel exchange. In conclusion we find that only a fraction of the 
polarizability of the nucleon is related to easily identified low-energy resonances  in the $s$-channel. The remaining 50\% or more are described by 
$t$-channel exchange of mesons. While we set out to investigate the structure of the nucleon, we find that we learn about the structure of pions and many-pion states !
\begin{table}[t]
%\begin{flushleft}
\caption{Contributions to the polarizabilities
from the dispersion integrals along the real axis and the half-circle, and the sum of both in units of $10^{-4} fm^{3}$. 
These results are compared to the predictions of relativistic $ChPT$ 
\protect\cite{Ber91} 
(results identical to the $\sigma$-model 
\protect\cite{Met96}) 
and to heavy baryon $ChPT$ 
\protect\cite{Ber94} 
as well as to the experimental data 
\protect\cite{Mac95}.
}
\begin{center}
\begin{tabular}{|c||ccc||ccc|}
%\hline
\hline
 & real axis &half-circle & sum & rel. $ChPT$ & $HB ChPT$ & 
data \\
\hline
$\alpha$ & 5 & 7 & 12 & 7.5 & 10.5 & 12.1 \\
\hline
$\beta$& 8 & -6 &2& -2.0 & 3.5 & 2.1 \\
\hline
\end{tabular}
\end{center}
\end{table}
\section{Calculations of Polarizabilities}

Following many investigations of polarizabilities in the framework of constituent and chiral quark as well as soliton models, the first prediction from $ChPT$ was made by Bernard et al \cite{Ber91}. As an example the electric polarizability of the proton is obtained in terms of the mass ratio
$\mu = m_{\pi}/m$,
\begin{equation}
\alpha_{p} = \frac{e^{2} g_{\pi N}^{2}}{192 \pi^{3} m^{3}} \left\{\frac{5\pi}{2\mu}
+ 18 \hbox{ln} \mu + \frac{33}{2} + [\mu])\right\}\,\,,
\end{equation}
Incidentally, this result is in complete agreement with the linear $\sigma$-model in the case of an infinite $\sigma$ mass \cite{Met96}.
As is to be expected, $\alpha_{p}$ diverges in the soft-pion limit $\mu\rightarrow 0$. In that limit the relation $\alpha_{p} = \alpha_{n} = 10\beta_{p} = 10\beta_{n}$ is obtained, in reasonable agreement with the data.  In order to 
obtain a more consistent expansion in terms of $m^{-1}$, the calculation has been repeated in heavy baryon $ChPT$ \cite{Ber94} up to $O(q^{4})$. The result agrees reasonably well with the experiment (see Table 2). 
However, there appear low energy constants (LEC) which 
have to be fixed in a model-dependent way by resonance saturation. Obviously
these constants describe or rather ''integrate out'' the physics at short distance. If the dispersion analysis of L'vov is correct and about half of the polarizability is given by high-energy contributions at $\omega > 1.5 GeV$, 
the influence of LEC's should be substantial and more-loop calculations will 
hardly improve the situation. 

\section{Generalized Polarizabilities}

Following the early work of Plazcek \cite{Pla34} for $RCS$, the cross section of $VCS$ may be expanded into a series of multipoles \cite{Are74}. The differential cross section 
is then obtained by replacing the density matrix for real photons with 
polarization in the initial state by the appropriate matrix for virtual photons, $\rho_{\lambda'\lambda}^{(i)}$.
This density matrix is related to the usual kinematical function in electron scattering, e.g. $\rho_{00} = V_{L}, \rho_{11} = V_{T}$ etc. Schematically the cross section for $VCS$ can be written in the form
\begin{equation}
\frac{d\sigma}{d\Omega_{e'} d\Omega_{\gamma '} d\omega} = \sum\limits_{LL' KK'j}
P_{j}^{\ast} (L'L, q' q) P_{j} (K' K, q' q)\\
h_{j} (L'L K' K, \rho^{(i)})\,\,,\nonumber
\end{equation}
with well-defined kinematical functions $h_j$. All the physics is contained in the polarizabilities
which are functions of the multipoles 
${\cal{M}}^{L\rho}$, with  $L$ the total angular momentum of the electromagnetic radiation and $\rho$ defining its polarization ($\rho = 0$ longitudinal, $\rho = 1$ magnetic, and $\rho = 2$ electric),
\begin{eqnarray}
P_{j} (L' \rho ', L\rho; q' q) \sim \sum\limits_{n} 
\left\{ \begin{array}{ccc}
 L'  & L &  j \\
 J_{i} & J_{f}  & J_{n} \\
\end{array}\right\}
< J_{f}\parallel {\cal{M}}^{L'\rho'} 
\parallel J_{n}>\\
(\omega_{n} - \omega - i\Gamma_{n}/2)^{-1} < J_{n} \parallel {\cal{M}}^
{L\rho}\parallel J_{i}> +\,\,u-\hbox{channel}\,\,. \nonumber
\end{eqnarray}
Note that in the general case of resonance scattering a width $\Gamma_n$ has to be included. Below threshold this width vanishes and the polarizabilities  are real functions. The number of  polarizabilities involved depends on the spin of the hadron. For elastic photon scattering off a scalar or pseudoscalar particle, e.g. a pion, $J_i = J_f = 0$, and the 6-j symbol in equation (9) allows only for $j = 0$, the scalar polarizabilities. If we further assume low-energy outgoing photons, we may use the dipole approximation , $L' = 1$. As a consequence the spin of the intermediate states is $J_n = 1$, and 
the virtual polarizabilities of a pion contain the combinations:
$E1 \times E1, E1 \times L1$ or $L1 \times L1$ (exciting $1^+$ states), and $M1 \times M1$ (exciting $1^-$ states).
In the case of a spin $1/2$ particle, $J_i = J_f = 1/2$, there are both scalar ($j = 0$) and vector ($j = 1$) polarizabilities leading to states with $J_n = 1/2$ or $3/2$. The higher the spin of the hadron, the more  polarizabilites can be reached, e.g. tensor ($j = 2$) polarizabilities in the case of a deuteron.
\begin{table}[t]
%\begin{flushleft}
\caption{Multipoles and generalized polarizabilities for the nucleon. Note 
that the electric multipoles $El$ are always expressed by $Ll$, and only the difference between the two is an independent quantity and given in the table by $\hat{P}$.}
%\end{flushleft}
%\vspace{0.4cm}
\begin{center}
\begin{tabular}{|l|c|l|l|}
%\hline
\hline
multipoles & S & intermed. states & ($\rho'L', (\rho) L)S$\\
\hline
$L1\times L1$& 0, 1 &$ 1/2^{-}, 3/2^{-}$ & $P (01, 01) S$ \\
$L1\times E1$& 0,1& $1/2^{-} , 3/2^{-}$ & $\hat{P}(01, 1)S$ \\
$M1\times M1$& 0, 1& $1/2^{+}, 3/2^{+}$&$ P(11, 11) S$\\
$L1 \times M2$ & 1 &$3/2^{-}$ &$P(01, 12) 1$\\
$M1 \times L2$& 1 & $3/2^{+}$&$P(11, 02)1$\\
$M1\times L0$ &1&$1/2^{+}$& $P (11,00) 1$\\
$M1\times E2$& 1& $1/2^{+}$ &$\hat{P} (11,2) 1$\\
\hline
\end{tabular}
\end{center}
\end{table}
It should be observed that the polarizabilities are functions of both $q$ and $q'$. In order to simplify this situation, Guichon \cite{Gui95} has introduced generalized polarizabilities ($GP$) by dividing 
out the low energy dependence of the multipoles $\sim(q')^{L'} q^{L}$, which is to be expected in the Siegert limit of vanishing 3-momenta. Since the outgoing photon is assumed to be soft, these $GP's$ are defined in the limit of $q'\rightarrow 0$,
\begin{equation}
P^{(\rho'L, \rho L)S} = \lim\limits_{q'\rightarrow 0} ((q')^{-L'}
q^{-L} P_{S} (L'\rho', L\rho ; q' q))\,\,.
\end{equation}    
Since the electric ($\rho = 2$) and longitudinal ($\rho = 0$) multipoles
are related by current conservation in the Siegert limit, the electric 
polarizability may be defined by the longitudinal current. The difference
between the transverse current and the longitudinal one, which is of relative 
$O (q^{2})$, is then expressed by additional ''mixed''
polarizabilities, $\hat{P}^{(\rho' L', L)S}$. Altogether this leads 
to 10 generalized polarizabilities to the order considered (see Table 3).
\begin{figure}
\centerline{\psfig{figure=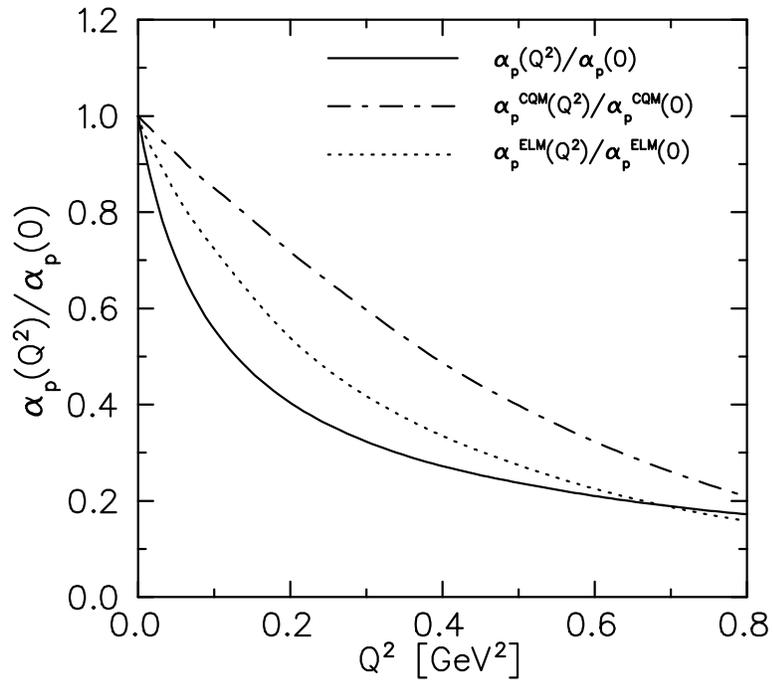,width=10cm}}
\caption{The generalized electric polarizability, normalized to its value at 
$Q^{2} = 0$, for the proton. The predictions are obtained from $CQM$ (Ref. 24),
$ELM$ (Ref. 25) and $LSM$ (Ref. 19) for the dashed-dotted and full curve, 
respectively.}
\end{figure}
The polarizabilities for $VCS$ off the nucleon have been calculated only recently. The pioneering work in this field was performed by Guichon et al.
\cite{Gui95,Liu96}, who investigated $VCS$ in the framework of the 
constituent quark model ($CQM$), taking account of nucleon and resonance intermediate states in the $s$- and $u$-channels. These polarizabilities have a  Gaussian dependence on momentum transfer, which is a reasonable description of the transition form factors for the low-lying resonances. A further approach has been worked out in the framework of an effective Lagrangian model ($ELM$) by Vanderhaeghen
\cite{Van96}, who included both the low-lying resonances  with effective couplings 
and $t$-channel exchange diagrams ($\pi^0$ and $\sigma$). Furthermore,
there will be reports at this meeting on calculations in the linear sigma model ($LSM$) \cite{Met96} and in the heavy baryon version of $ChPT$ \cite{Hem96}. 
The $LSM$ is a convenient toy model to study the general structure of $VCS$, because it fulfills all
the relevant symmetries (Lorentz, gauge and chiral invariance). Since it is
essentially free of parameters (at least in the limit of $m_{\sigma}\rightarrow \infty$), it cannot give a realistic description of the nucleon but may serve as a check of general properties the amplitudes should
obey. As an example such calculations have established interesting relations between the $GP's$, which follow from general arguments but are difficult to 
obtain in phenomenological or nonrelativistic theories. The approach of $HB ChPT$, on the other side, describes the most 
general Lagrangian compatible with $QCD$. Therefore, it should reproduce the 
experiment provided that the loop-expansion can be carried out to sufficient order and at the expense of introducing more and more $LEC$'s.

Figure 3 shows a strong model-dependence of the predictions even at 
$Q^{2}\approx 0.2 GeV^{2}$. The ''pion cloud'' contributions of the $LSM$ decrease much faster as function of $Q^2$ than in models with Gaussian or dipole form factors. However, there remain ''hard'' components probably 
due to contact interactions, and the polarizabilities as function of $Q^2$ 
level off as $Q^2$ approaches  1 $GeV^{2}$ and become comparable with the other models.

At the higher energies $VCS$ has been studied in the diquark model by Kroll et al \cite{Kro95}. This model gives a good description of the nucleon at energies of a few $GeV$. It will provide a good basis for the analysis of $VCS$ in the multi-$GeV$ region, with the aim of studying the wave function of the nucleon.

\section{Outlook}

Undoubtedly the potential of $VCS$ for structure investigations of the nucleon is quite considerable, but its experimental realization is very difficult 
due to a large background of the Bethe-Heitler (BH) process (see Figure 2). 
These two reactions add coherently,
and due to the small mass of the electron, BH radiation leads to large cross sections in a cone about the directions of the incoming and outgoing electrons. 
In an intermediate region information may be obtained from the interference term. Being linear in the polarizabilities, it enhances the small effects expected
from $VCS$. Since it is not symmetrical about the axis of the virtual photon, its  structure is completely different
from the usual response functions in electroproduction. As has been shown  for typical MAMI energies, the polarizability shows up only at angles far away from the dominating cones of BH
radiation, i.e. at backward angles and/or for noncoplanar  geometry. Altogether the polarizabilities give rise to a 10 - 20 \% effect at energies of the order of 1 $GeV$, in a kinematical region with differential cross sections of the order of $0.1 pb/MeVsr^2$. At the higher energies of CEBAF, $VCS$ is expected  to dominate at backward and non-coplanar angles, with expected differential cross sections of the order of $1 pb/GeV sr^2$.

The first results of the A1 collaboration at MAMI will be presented in this workshop \cite{And95,Roc96}. There will also be a report on higher order radiative corrections, which have been shown to be of
similar importance as the $GP$'s one hopes to extract from the data \cite{Mar96}.

Another $VCS$ experiment has been proposed for CEBAF\cite{And93} at higher energies. Its aim will be to study $GP$'s, the resonance structure of $VCS$ and the asymptotics
at large energy. The experiment has been rated highly ($A^{-}$) by the CEBAF
PAC, and has been scheduled for 1997. In a further experiment it is planned to explore $VCS$ at large $Q^2$ in order to study the parton wave functions \cite{Bre94}. This proposal was conditionally approved, on condition that the previous experiment will be successful. 

There are further proposals for the project ELFE \cite{Fon93} in the region of $s\sim 10 GeV^2$. As has been pointed out, the differential cross section is 
expected to scale in this region with $s^{-6}$. Calculations by Kroll et al.
\cite{Kro95} have shown that the process will be dominated by the transverse structure function, leading to $d\sigma_{T}/dt \sim nb/GeV^2$. The ultimate aim of this difficult experiment is to study the wave function of the valence quarks in the nucleon, assuming that higher Fock states can be neglected or calculated sufficiently well. 

In conclusion $VCS$ is a great challenge for the international collaborations which have been formed to study this kind of physics. Such activities have been
motivated by the high present-day interest in $QCD$-based descriptions of the 
hadrons. The experiments may be tough, but they certainly promise to give new and independent information on the structure of the nucleon.

\end{document}